# Aspects of the Synthesis of Thin Film Superconducting Infinite-Layer Nickelates

Kyuho Lee[1,2,a], Berit H. Goodge[3], Danfeng Li[1,4,b], Motoki Osada[1,5], Bai Yang Wang[1,2], Yi Cui[1,5], Lena F. Kourkoutis[3,6] & Harold Y. Hwang[1,4]

[1]*Stanford Institute for Materials and Energy Sciences, SLAC National Accelerator Laboratory, Menlo Park, CA 94025, USA*
[2]*Department of Physics, Stanford University, Stanford, CA 94305, USA*
[3]*School of Applied and Engineering Physics, Cornell University, Ithaca, New York 14853, USA*
[4]*Department of Applied Physics, Stanford University, Stanford, CA 94305, USA*
[5]*Department of Materials Science and Engineering, Stanford University, Stanford, CA 94305, USA*
[6]*Kavli Institute at Cornell for Nanoscale Science, Ithaca, New York 14853, USA*

**Abstract**

The recent observation of superconductivity in $Nd_{0.8}Sr_{0.2}NiO_2$ calls for further investigation and optimization of the synthesis of this infinite-layer nickelate structure. Here, we present our current understanding of important aspects of the growth of the parent perovskite compound via pulsed laser deposition on $SrTiO_3$ (001) substrates, and the subsequent topotactic reduction. We find that to achieve single-crystalline, single-phase superconducting $Nd_{0.8}Sr_{0.2}NiO_2$, it is essential that the precursor perovskite $Nd_{0.8}Sr_{0.2}NiO_3$ thin film is stabilized with no visible impurity phases; in particular, a Ruddlesden-Popper-type secondary phase is often observed. We have further investigated the evolution of the soft-chemistry topotactic reduction conditions to realize full transformation to the infinite-layer structure with no film decomposition or formation of other phases. We find that capping the nickelate film with a subsequent $SrTiO_3$ layer provides an epitaxial template to the top region of the nickelate film, much like the substrate. Thus, for currently optimized growth conditions, we can stabilize superconducting single-phase $Nd_{0.8}Sr_{0.2}NiO_2$ (001) epitaxial thin films up to ~ 10 nm.

___________________________
[a]kyuho@stanford.edu
[b]denverli@stanford.edu

**Introduction**

Low-temperature oxygen deintercalation of the Ruddlesden-Popper (RP) series of nickelates $Ln_{n+1}Ni_nO_{3n+1}$ (Ln = lanthanides) gives rise to $Ln_{n+1}Ni_nO_{2n+2}$ structures with layered $NiO_2$ square planes and formal nickel oxidation state of $Ni^{1+1/n}$.[1-7] Notably, an unusual form of $Ni^+$ is reached in the $n = \infty$ infinite-layer nickelate $LnNiO_2$, realizing possible structural and electronic analogs to the undoped parent compound of layered cuprate high-temperature superconductors.[8-12] The synthesis of these infinite-layer nickelates was first reported in 1983, where polycrystalline perovskite $LaNiO_3$ was reduced to $LaNiO_2$ with hydrogen gas as the reducing agent.[1,2] It was later shown in 1999 and onwards that this topotactic reduction process can be achieved more reproducibly at lower temperature by using metal hydrides for reduction.[4,5,13] This technique was further extended to epitaxial nickelate thin films, with the first demonstration in 2009 for $LaNiO_3$ (001) epitaxial thin films using $CaH_2$ as the reducing agent.[13,14]

Motivated to explore the analogy to superconducting cuprates, we have recently observed superconductivity in chemically doped $Nd_{0.8}Sr_{0.2}NiO_2$ (001) epitaxial thin films grown on $SrTiO_3$ (001) substrate by pulsed laser deposition (PLD).[15] This finding warrants the systematic investigation of its superconducting and normal state properties, for which establishing a reproducible synthetic route is critical. There are two key issues in stabilizing $Nd_{0.8}Sr_{0.2}NiO_2$ (001) epitaxial thin films. First is the instability of the precursor perovskite phase. While chemical doping by strontium brings the nickel oxidation state of the infinite-layer phase (nominally $Ni^{1.2+}$) closer to the thermodynamically stable $Ni^{2+}$, it results in a rather extreme formal nickel oxidation state of $Ni^{3.2+}$ in the $Nd_{0.8}Sr_{0.2}NiO_3$ perovskite precursor. This chemical instability adds significantly to the existing synthesis challenges of the undoped perovskite $NdNiO_3$, namely the nontrivial fluctuation of the film quality upon subtle changes in growth and post-annealing conditions.[16-19] In addition, tailoring the substrate choice to minimize lattice mismatch with the infinite-layer phase (–0.4% for the $SrTiO_3$ (001) substrate)[5] forces a large tensile strain (+2.6% with $SrTiO_3$ (001))[20] on the perovskite nickelate. These factors pose an interesting materials challenge to forming the aimed infinite-layer structure, the crystallographic quality of which is found to be heavily dependent on that of the precursor perovskite structure.[4]

Second, previous studies have shown that it is difficult to stabilize uniform, single-crystalline infinite-layer nickelate films from soft-chemistry topotactic reduction of the perovskite.[13, 21-23] For example, reduction studies on LaNiO$_3$ have shown that, besides the infinite-layer LaNiO$_2$ (001), phases such as brownmillerite LaNiO$_{2.5}$ and *a*-axis oriented LaNiO$_2$ (100) can appear during reduction.[13, 21] A previous study on NdNiO$_3$ reduction also indicated that a fluorite defect phase can be introduced on top of the infinite-layer NdNiO$_2$ (001) films under certain annealing conditions.[22] Depending on the reduction conditions, decomposition of the infinite-layer phase at the upper region of the film was also observed.[23] These results indicate the need of careful optimization of the reduction conditions, and, perhaps, adjustments in the structural design of the film to promote single-phase stabilization.

In this study, we survey the stabilization of single-phase, single-crystalline Nd$_{0.8}$Sr$_{0.2}$NiO$_2$ (001). In the first section of this paper, we examine the optimization of PLD growth of perovskite Nd$_{0.8}$Sr$_{0.2}$NiO$_3$ (001) on SrTiO$_3$ (001) substrate. We discuss two different optimized growth conditions for Nd$_{0.8}$Sr$_{0.2}$NiO$_3$, which are based on two different laser fluences. In the second section of this paper, we present studies on the CaH$_2$-assisted topotactic reduction of the Nd$_{0.8}$Sr$_{0.2}$NiO$_3$ (001) precursor phase, discussing the effect of a SrTiO$_3$ capping layer on the reduction process and the evolution of the nickelate film as a function of reduction time. Finally, we discuss how the choice of the growth conditions affects the crystallinity and the superconducting transition of Nd$_{0.8}$Sr$_{0.2}$NiO$_2$ (001).

5 × 5 mm$^2$ TiO$_2$-terminated SrTiO$_3$ (001) substrates were pre-annealed for 30 minutes at 930 °C under oxygen partial pressure $P_{O_2} = 5 \times 10^{-6}$ Torr to achieve a sharp step-and-terrace surface prior to film growth. Undoped and Sr-doped nickelate films were grown on these substrates by PLD with a KrF excimer laser ($\lambda$ = 248 nm), using mixed-phase polycrystalline targets of Nd$_2$NiO$_4$ + NiO and mixed-phase polycrystalline targets of (Nd$_{0.8}$Sr$_{0.2}$)$_2$NiO$_4$ + NiO, as confirmed by powder x-ray diffraction (XRD), respectively. In this study, we used 1.25 mm × 1.90 mm and 1.60 mm × 2.45 mm uniform rectangular laser spots for ablation, which we denote as the small and large laser spots, respectively. These laser spots were formed by imaging an aperture. The targets were prepared by sintering mixtures of stoichiometric amounts of Nd$_2$O$_3$, SrCO$_3$, and NiO powders at 1350 °C for 12 hours, with two extensive intermediate grinding (resulting in a fine powder) and pelletizing steps after initial decarbonation at 1200 °C for 12 hours. This process leads to a uniform polycrystalline target (density of ~ 75%) with no visible grains embedded, indicating a grain size conservatively below 2 microns. Details on the PLD growth conditions of SrIrO$_3$ (001) epitaxial films (Figure 7(a)) can be found in Ref. 24.

The deposited films were cut into two pieces of size 2.5 × 5 mm$^2$. Loosely wrapped in aluminum foil to avoid direct contact with the reducing agent, each piece was vacuum-sealed (pressure < 0.1 mTorr) with 0.1 g of CaH$_2$ powder in a Pyrex glass tube. The tube was then heated to the desired temperature to perform reduction. The temperature ramping rate was fixed at 10 °C min$^{-1}$.

XRD symmetric $\theta$-$2\theta$ scans of the nickelate films were measured using a monochromated Cu $K_{\alpha 1}$ ($\lambda$ = 1.5406 Å) source. Temperature-dependent resistivity ($\rho(T)$) measurements were conducted in a four-point geometry using aluminum wire-bonded contacts. In some cases, gold contact pads were evaporated using electron-beam evaporation before wire-bonded contacts were made. Cross-sectional scanning transmission electron microscopy (STEM) specimens were prepared using a standard focused ion beam (FIB) lift-out process on a Thermo Scientific Helios G4 X FIB. High-angle annular dark-field STEM (HAADF-STEM) images were acquired on an aberration-corrected FEI Titan Themis at 300 keV with probe convergence semi-angles of 21 - 30 mrad and inner and outer collection angles of 68 and 340 mrad, respectively.

## I. Stabilizing Perovskite Nd$_{0.8}$Sr$_{0.2}$NiO$_3$ (001)

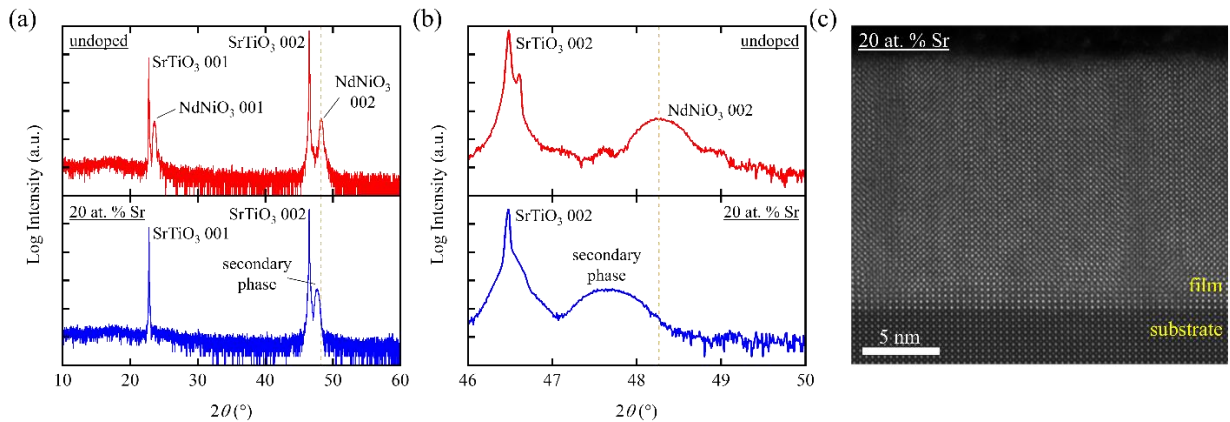

FIG. 1. (a) XRD symmetric $\theta$-$2\theta$ scans of a NdNiO$_3$ film (top) and the film grown using the 20 at. % Sr-doped target under the same growth conditions (bottom). Both films are grown on SrTiO$_3$ (001) substrates, with film thickness of ~ 20 nm. The dotted line indicates the NdNiO$_3$ 002 peak position. (b) Expanded view of panel (a) near the SrTiO$_3$ 002 peak. (c) Cross-sectional HAADF STEM image of the 'secondary phase' film.

Literature on PLD-growth of undoped NdNiO$_3$ reports a relatively flexible range of growth conditions, with substrate temperature $T_s$ ranging from 600 to 750 °C, $P_{O_2}$ ranging from 100 to 200 mTorr, laser fluence $F$ ranging from 1.5 to 2.1 J cm$^{-2}$, and laser frequency $f$ ranging from 2 to 30 Hz.[15-17, 19, 25, 26] Figure 1 shows the XRD symmetric $\theta$-$2\theta$ scan of a ~ 20 nm NdNiO$_3$ (001) film grown by PLD on single-crystal SrTiO$_3$ (001) substrate under the growth conditions of $T_s = 600$ °C, $P_{O_2} = 200$ mTorr, $F = 1.4$ J cm$^{-2}$, and $f = 4$ Hz, using the small laser spot. Considering the tensile strain induced by the substrate, the extracted film $c$-lattice constant of 3.77 Å is in good agreement with the pseudocubic bulk lattice constant of NdNiO$_3$ (3.807 Å).[27] The prominent NdNiO$_3$ 001 peak and the presence of fringes around the film peaks in the symmetric $\theta$-$2\theta$ scan (Figures 1(a) and 1(b)) are indications that the growth conditions are well within the optimal growth window of NdNiO$_3$.[28]

In many material systems, it is often the case that doping or partial cationic substitution requires minimal or no change in the PLD growth conditions.[29-33] However, when the identical growth conditions above are employed using the 20 at. % Sr-doped target, the resultant film symmetric $\theta$-$2\theta$ scan is far from that of highly-crystalline Nd$_{0.8}$Sr$_{0.2}$NiO$_3$ (001) (Figures 1(a) and 1(b)). Namely, the perovskite 001 peak is absent (Figure 1(a)), and the extracted $c$-lattice constant of 3.80 Å is nontrivially larger than that of the optimized NdNiO$_3$ (001) film (Figure 1(b)). Both of these features have been previously observed in undoped nickelate films.[17, 34] The cross-sectional HAADF STEM image of this film reveals that it is densely populated with vertical RP-type faults (Figure 1(c)). These defects are formed when an AO rocksalt layer (where A corresponds to the A-site cation) stabilizes in between the perovskite layers.[34-36] The frequent inclusion of these rocksalt layers breaks the structural long-range order of the perovskite phase and makes the film analogous to a highly disordered sequence of in-plane oriented RP phases. Perhaps surprisingly, the observed diffraction pattern of the secondary phase film matches well to the (110) oriented trilayer RP phase (Nd$_{0.8}$Sr$_{0.2}$)$_4$Ni$_3$O$_{10}$, which has a 220 diffraction peak aligning well to the observed film peak at $2\theta \approx 47.84°$.[37] Also, this phase has no 110 peak by symmetry.[37] While this would be consistent with the absence of a film peak near the SrTiO$_3$ 001 substrate peak, the presence of the visible stacking disorder should tend to diminish the strict extinction of this peak. As will be later shown, this phase with densely populated vertical RP-type faults behaves very differently from the perovskite in terms of topotactic reduction and transport properties. Given the high degree of disorder and structural ambiguity, for simplicity we denote this as a 'secondary phase'. Overall, these observations indicate that Sr doping significantly reduces the growth window of the perovskite phase, and that further optimization of growth conditions is required.

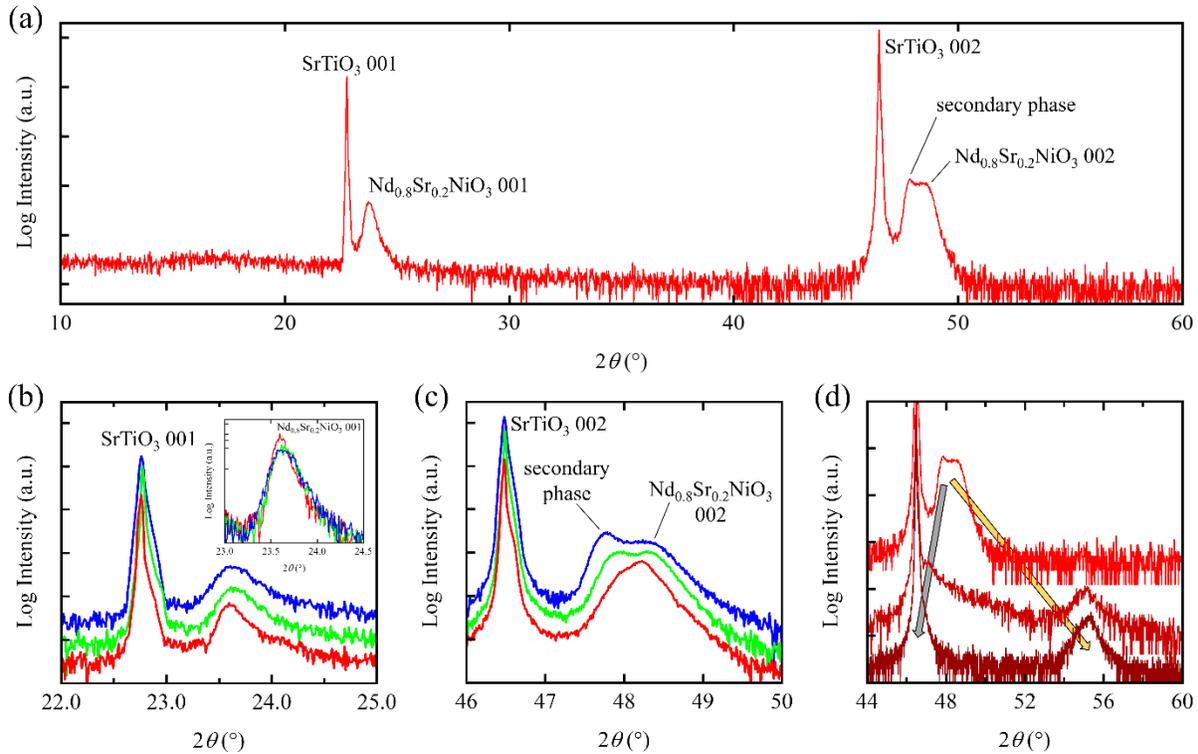

FIG. 2. (a) XRD symmetric $\theta$-$2\theta$ scan of a ~ 60 nm film grown on SrTiO$_3$ (001) using the 20 at. % Sr-doped target under partially optimized growth conditions. (b) XRD symmetric $\theta$-$2\theta$ scans of three ~ 60 nm thick doped films grown consecutively (bottom to top) on SrTiO$_3$ (001), magnified near the SrTiO$_3$ 001 peak. The curves are vertically offset for clarity. The inset shows the scans without vertical offset, showing the systematic broadening of the 001 film peak over target ablation. (c) Same scans of panel (b) magnified near the SrTiO$_3$ 002 peak. (d) XRD symmetric $\theta$-$2\theta$ scans of the film shown in panel (a) at different stages of the reduction process; as-grown (top), intermediate reduction (middle; 240 °C, 1 hour), and after complete reduction (bottom; 240 °C, 2 hours). The two arrows indicate the evolution of the secondary-phase peak (left) and the perovskite-phase peak (right). The curves are vertically offset for clarity.

At an intermediate stage of our attempts to optimize the growth conditions, we observed a co-stabilization of the perovskite phase and the secondary phase. Figure 2 shows the structural characteristics of samples (~ 60 nm in thickness) grown under two nominally similar growth conditions: 1) $T_s$ = 600 °C, $P_{O_2}$ = 35 mTorr, $F$ = 0.9 J cm$^{-2}$ and $f$ = 2 Hz with the large laser spot (Figures 2(a) and 2(d)), and; 2) $T_s$ = 600 °C, $P_{O_2}$ = 70 mTorr, $F$ = 0.9 J cm$^{-2}$ and $f$ = 4 Hz with the large laser spot (Figure 2(b) and 2(c)). All of these samples show the Nd$_{0.8}$Sr$_{0.2}$NiO$_3$ 00$l$ peaks superposed with the secondary phase peak, which has a smaller $2\theta$ value than the Nd$_{0.8}$Sr$_{0.2}$NiO$_3$ 002 peak in the XRD symmetric $\theta$-$2\theta$ scan (Figures 2(a) and 2(c)). In addition, we find that the population of the two observed phases changes as a function of target history. The XRD symmetric $\theta$-$2\theta$ scan of three samples grown consecutively under fixed growth conditions (Figures 2(b) and 2(c)) shows that the secondary phase gradually dominates over the perovskite phase with increasing target ablation. This is in line with the previous observation of limited film reproducibility and nickel enrichment of the target over time in the PLD study of NdNiO$_3$;[19] the favored ablation of the A-site cations from the target may increasingly promote the stabilization of the A-site-rich secondary phase over time. To avoid ambiguities arising from target history, we subsequently re-polished the target surface after each film growth.

It is interesting to see how these partially optimized mixed-phase films transform upon CaH$_2$-assisted topotactic reduction. Figure 2(d) shows the evolution of the XRD $\theta$-$2\theta$ peaks of a mixed-phase film over the reduction process. The right of the as-grown double peak, which corresponds to the perovskite phase, shifts further rightward upon reduction and saturates at a $2\theta$ value corresponding to $c$ = 3.32 Å, indicating the successful transformation of the perovskite phase to the infinite-layer structure.[5, 13] In contrast, the left of the as-grown double peak shifts further leftward towards the SrTiO$_3$ 002 peak position. Again assuming the secondary phase can be approximately described as (Nd$_{0.8}$Sr$_{0.2}$)$_4$Ni$_3$O$_{10}$ (110), the corresponding reduced structure (Nd$_{0.8}$Sr$_{0.2}$)$_4$Ni$_3$O$_8$ (100) will have a 200 peak near $2\theta \approx$ 46.32°, which is very close to the SrTiO$_3$ 002 peak at $2\theta \approx$ 46.47°.[3] Note that the change in the crystallographic notation is due to the difference in the conventional unit cell space group of (Nd$_{0.8}$Sr$_{0.2}$)$_4$Ni$_3$O$_{10}$ ($P2_1/a$) and (Nd$_{0.8}$Sr$_{0.2}$)$_4$Ni$_3$O$_8$ ($I4/mmm$).[3, 37] This reduced structure also has no lower-order peak (i.e. 100 peak) by symmetry.[3]

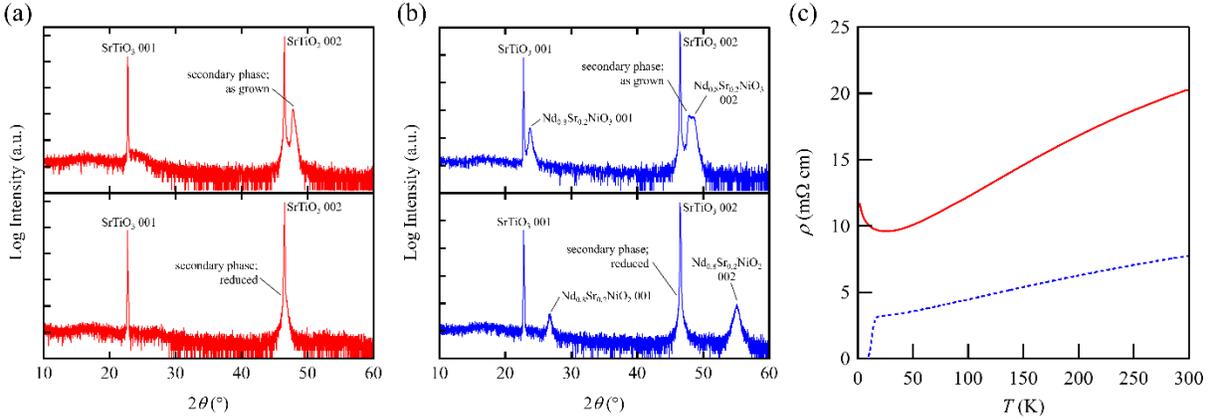

FIG. 3. (a) XRD symmetric $\theta$-$2\theta$ scans of the film which dominantly consists of the secondary phase (~ 40 nm in film thickness) before (top) and after (bottom) reduction (240 °C, 2.5 hours). (b) XRD symmetric $\theta$-$2\theta$ scans of the mixed-phase film before (top; same as Figure 2(a)) and after (bottom) reduction (240 °C, 2 hours). (c) $\rho$-$T$ curves of the film in panel (a) (solid curve) and panel (b) (dashed curve).

Such clear difference in the structural evolution of the two phases upon reduction translates to the transport properties of the two phases. For the film that dominantly consists of the secondary phase, with essentially no sign of the infinite-layer phase after reduction (Figure 3(a)), no evidence for superconductivity is found down to 2 K. For the mixed-phase film after reduction (Figure 3(b)) a superconducting transition is observed, with an onset at 14.7 K (point of maximum curvature), a midpoint at 12.6 K, and zero resistance at 7.2 K (indistinguishable from the noise floor) (Figure 3(c)). These observations indicate that the infinite-layer nickelate phase, not the reduced secondary phase, is superconducting. We emphasize that the presence of the perovskite 001 film peak, and the 002 film peak position, are the two strongest and most useful functional indicators for superconductivity; when the 001 peak is not observed and/or the 002 peak $2\theta$ position is below ~ 48°, the subsequently reduced film never exhibits superconductivity.

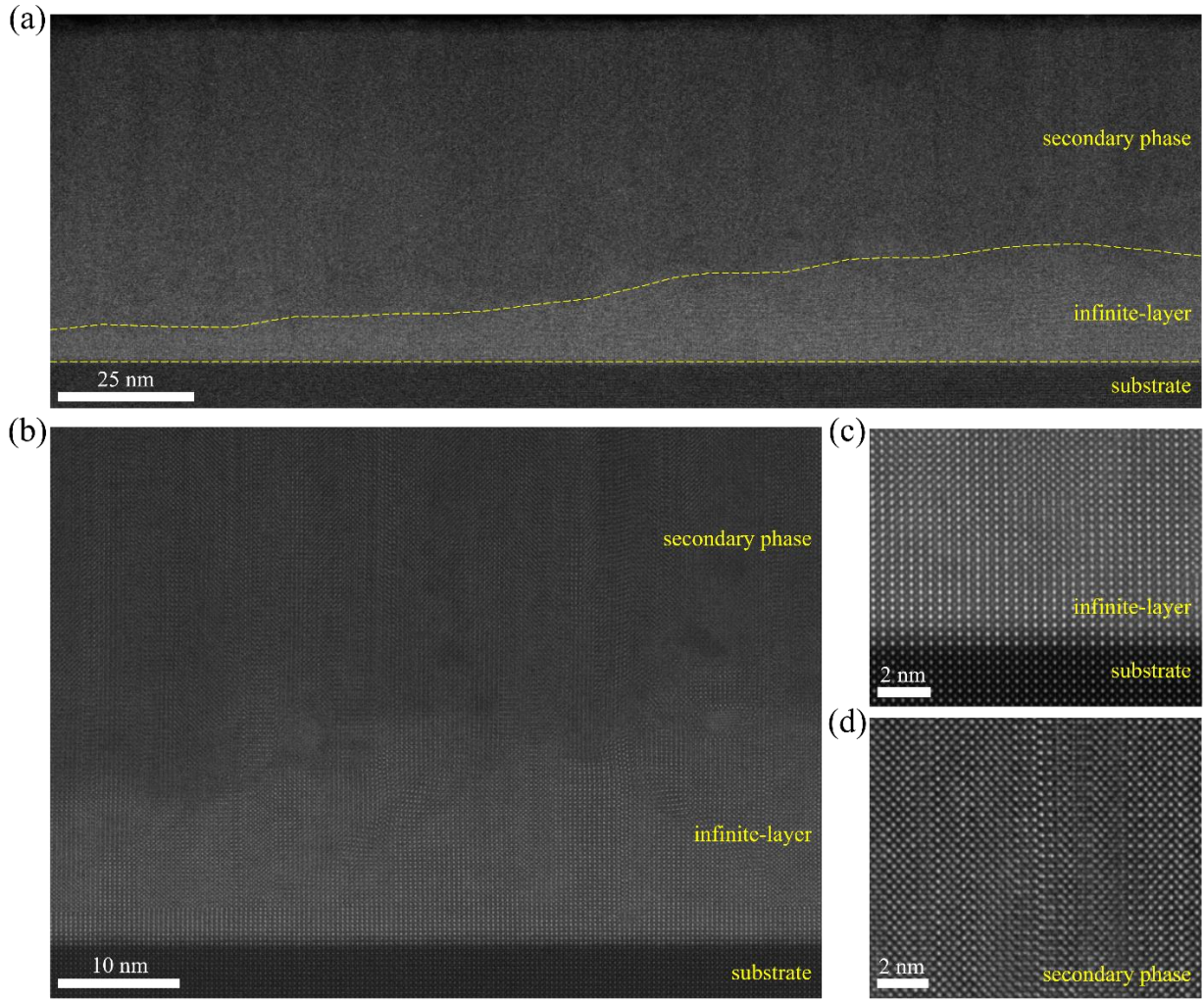

FIG. 4. (a) Cross-sectional HAADF STEM image of the mixed-phase film in Figure 2(a) after reduction. (b) Magnified image of panel (a). (c) Magnified image of panel (b) in the infinite-layer region. (d) Magnified image of panel (b) in the secondary phase region.

The cross-sectional STEM images of the reduced mixed-phase film (Figure 4) show a segregation of the two competing phases, where the infinite-layer phase is stabilized in the vicinity of the substrate and the secondary phase sits above the infinite-layer phase. Such preferred stabilization of the infinite-layer structure near the substrate has been observed in previous nickelate reduction studies.[22, 23] In particular, this was also observed for films grown by metal organic decomposition,[23] suggesting that the target history effects in the PLD growth of nickelates is not the primary factor for this phenomenon. Rather, this suggests that the epitaxial strain energy provided by the substrate plays an important role in stabilizing the perovskite phase during growth and the infinite-layer phase during the reduction process. Hence, growing thinner films can promote single-phase stabilization of the desired phase.

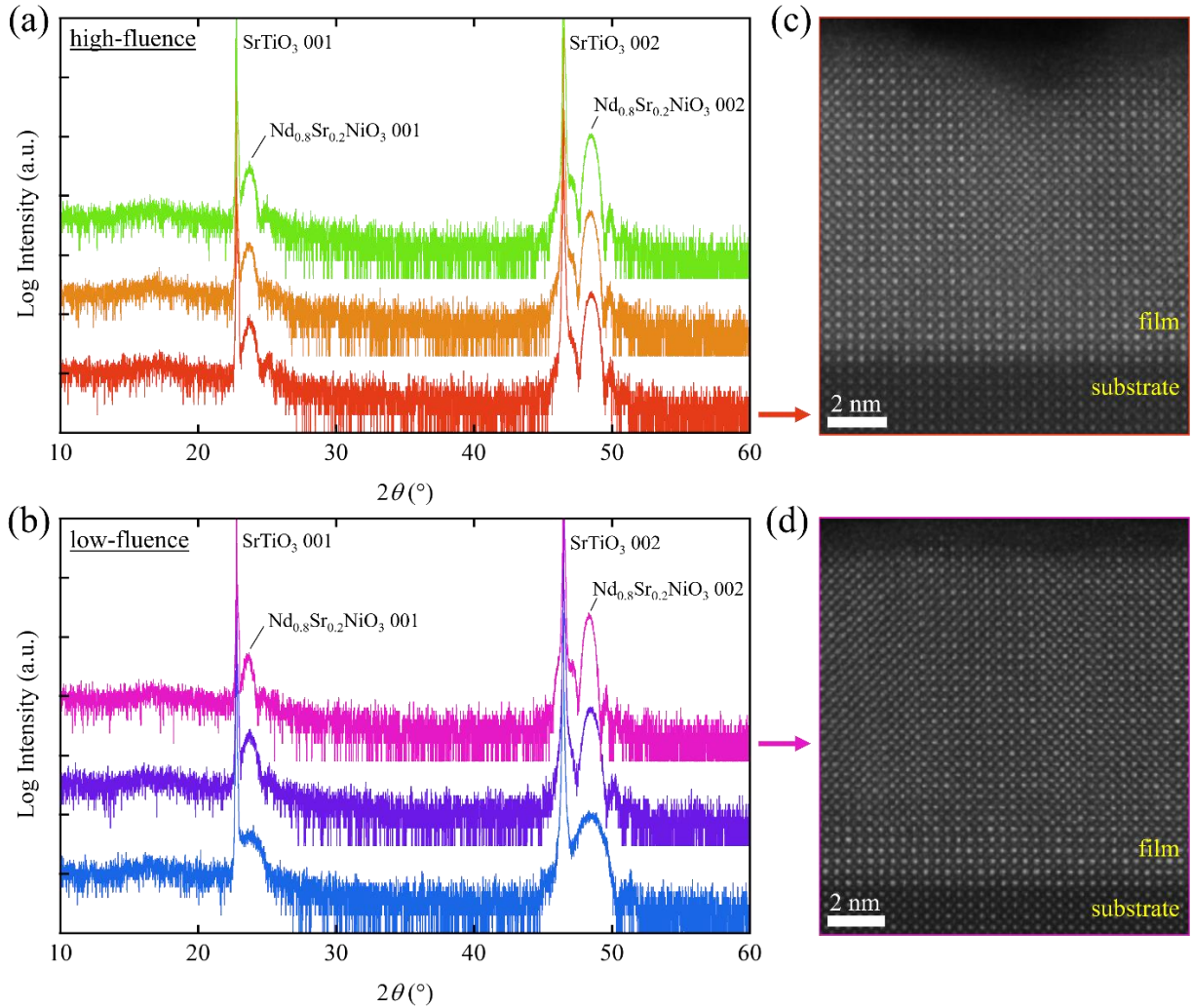

FIG. 5. (a) XRD symmetric $\theta$-$2\theta$ scans of three different $Nd_{0.8}Sr_{0.2}NiO_3$ films (from bottom to top: 10 nm, 10 nm, and 11 nm in film thickness) grown on $SrTiO_3$ (001) substrates under the optimized high-fluence growth conditions. The curves are vertically offset for clarity. (b) XRD symmetric $\theta$-$2\theta$ scans of three different $Nd_{0.8}Sr_{0.2}NiO_3$ films (from bottom to top: 5 nm, 9 nm, and 12 nm in film thickness) grown on $SrTiO_3$ (001) substrates under the optimized low-fluence growth conditions. The curves are vertically offset for clarity. (c) Cross-sectional HAADF STEM image of the film corresponding to the bottom scan in panel (a). (d) Cross-sectional HAADF STEM image of the film corresponding to the top scan in panel (b).

By further empirically optimizing the growth conditions and keeping the film thickness below ~ 15 nm, we were able to obtain $Nd_{0.8}Sr_{0.2}NiO_3$ (001) epitaxial films on $SrTiO_3$ (001) substrates with no visible secondary phase peaks in XRD under two different growth conditions. The first we denote as the 'high-fluence' growth conditions, with $T_s = 600$ °C, $P_{O_2} = 150$ mTorr, $F = 2.0$ J cm$^{-2}$, and $f = 4$ Hz using the small laser spot, while the second is in 'low-fluence' growth conditions, with $T_s = 600$ °C, $P_{O_2} = 70$ mTorr, $F = 1.0$ J cm$^{-2}$, and $f = 4$ Hz using the large laser spot. Figure 5 shows the XRD symmetric $\theta$-$2\theta$ scans of six optimized samples with film thickness ranging from 5 to 15 nm in these two growth conditions. All samples show $Nd_{0.8}Sr_{0.2}NiO_3$ 00$l$ film peaks with prominent 001 peak intensity and clean single 002 film peaks (Figures 5(a) and 5(b)). While vertical RP-type faults still exist (Figures 5(c) and 5(d)), the density of these defects is much lower than in the secondary phase. These observations indicate that single-phase $Nd_{0.8}Sr_{0.2}NiO_3$ (001) films with a low density of RP-type faults can be synthesized with the two above growth conditions in a reproducible fashion. While the films grown under these two conditions do show notable differences in the uniformity and crystallinity of the reduced films (see Section III), in both cases superconducting samples could be reproducibly achieved via topotactic reduction.

## II. Optimizing the Reduction Process for $Nd_{0.8}Sr_{0.2}NiO_2$ (001) Stabilization

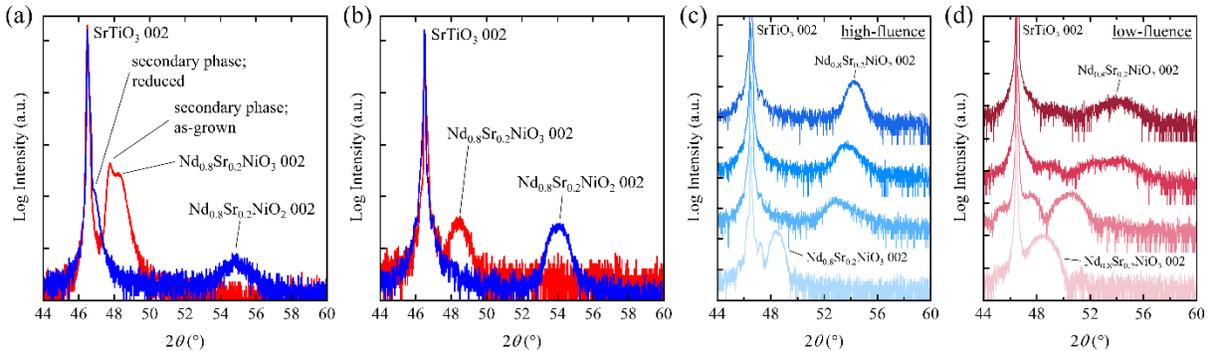

FIG. 6. (a) XRD symmetric $\theta$-$2\theta$ scan of a partially optimized sample (same growth conditions as films in Figure 2(b)) with film thickness of ~ 60 nm and no $SrTiO_3$ capping layer before (red) and after (blue) reduction (240 °C, 5 hours). (b) XRD symmetric $\theta$-$2\theta$ scan of a capped sample grown under the high-fluence conditions with film thickness of ~ 11 nm and cap thickness of ~ 25 nm before (red) and after (blue) reduction (9 hours at 260 °C, followed by 3 hours at 280 °C). (c) Evolution of $Nd_{0.8}Sr_{0.2}NiO_x$ 002 peak of a high-fluence capped sample with film thickness of ~ 11 nm and cap thickness of ~ 25 nm during the reduction process (from bottom to top: as-grown, 4

hours at 260 °C, additional 3 hours at 260 °C, additional 6 hours at 280 °C). (d) Evolution of $Nd_{0.8}Sr_{0.2}NiO_x$ 002 peak of a low-fluence sample with film thickness of ~ 5 nm and 5 unit cells of $SrTiO_3$ (001) capping layer during the reduction process (from bottom to top: as-grown, 0.5 hours at 240 °C, additional 1 hour at 240 °C, additional 1 hour at 240 °C).

During our soft-chemistry topotactic reduction experiments on the partially optimized films, we found the same challenges of film degradation that were observed in previous studies of undoped nickelates.[22, 23] Namely, only a portion of the perovskite film is converted to the infinite-layer structure, which is identified from the significantly reduced film peak intensity in the XRD symmetric $\theta$-$2\theta$ scan after the reduction process (Figure 6(a)).

There are several potential factors which can contribute to film degradation during reduction. If the reduction temperature $T_r$ is too high, the films can degrade before successfully forming the infinite-layer structure.[4, 5] This has been observed in the previous reduction study of undoped polycrystalline $NdNiO_3$ samples, where decomposition to $Nd_2O_3$ and Ni occurred when $T_r$ higher than 200 °C was employed with NaH as the reducing agent.[5] It is also possible that the infinite-layer phase is not accessible regardless of the value of $T_r$ because the reducing agent is not reactive enough; such is the case for the reduction of $NdNiO_3$ with hydrogen gas.[5] Therefore, the choice of an appropriate reducing agent along with careful optimization of $T_r$ and reduction time are required to achieve the highest crystallinity infinite-layer phase. We again note the structural support at the boundaries of the film. While epitaxial strain and structural support is provided by the substrate at the bottom of the film, promoting the infinite-layer phase, this is not the case for the top of the film away from the interface, which can lead to partial film degradation and the formation of impurity phases.[22, 23]

These factors suggest that capping the perovskite $Nd_{0.8}Sr_{0.2}NiO_3$ (001) film with $SrTiO_3$ may be helpful for the topotactic reduction to the infinite-layer structure in various ways. The capping layer can act as a protective barrier to prevent direct exposure of the film to the reducing agent, thus minimizing film decomposition. It can also act as a diffusion barrier, biasing the oxygen deintercalation to the in-plane direction and stabilizing (001)-oriented $Nd_{0.8}Sr_{0.2}NiO_2$. More generally, the epitaxial growth of $SrTiO_3$ (001) on top of the film provides the stabilizing proximity effect of the substrate on the top surface as well.

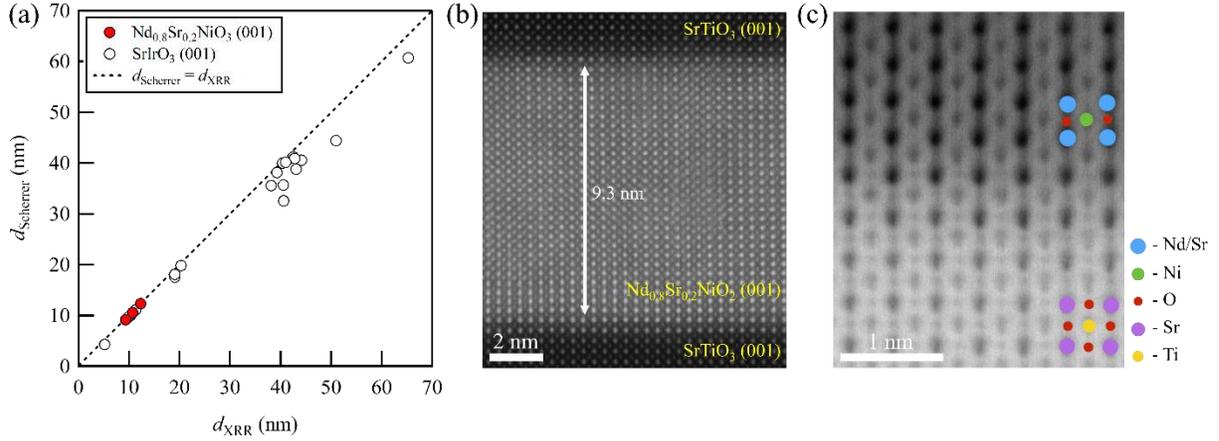

FIG. 7. (a) Film thickness values obtained from the Scherrer equation with $K = K_{film}$ ($d_{Scherrer}$) plotted against film thickness values obtained from x-ray reflectivity ($d_{XRR}$). (b) HAADF-STEM image of the capped $Nd_{0.8}Sr_{0.2}NiO_2$ film in Figure 6(b) after reduction. (c) Annular bright field (ABF) STEM image of the film in panel (b) with colored indication of the different atoms.

With these considerations in mind, we grew a $SrTiO_3$ (001) capping layer epitaxially on the $Nd_{0.8}Sr_{0.2}NiO_3$ (001) film under the same $T_s$ and $P_{O_2}$ as during the nickelate film growth, keeping the film thickness below ~ 15 nm. Indeed, the XRD $\theta$-$2\theta$ film peak intensity of the capped sample after reduction is much more prominent than that of the uncapped sample (Figure 6); in fact, the reduced film peak intensity is almost comparable to that of the as-grown film (Figures 6(b) - 6(d)). A direct quantitative measure of how much of the film has reduced to the infinite-layer structure is the comparison between the total thickness of the perovskite phase in the as-grown film and the total thickness of the infinite-layer phase in the reduced film. X-ray reflectivity (XRR) is a standard *ex situ* measurement technique for obtaining film thickness.[38] However, due to the small electron density contrast between the infinite-layer phase and the secondary phase, the XRR measurements alone are unable to provide a good estimate of the infinite-layer phase thickness.[38] In addition, the presence of the $SrTiO_3$ (001) capping layer complicates the thickness extraction from XRR. Instead, as an approximate measure, the Scherrer equation

$$d_{\text{Scherrer}} = \frac{K\lambda}{b \cos(\theta)}, \qquad (1)$$

where $K$ is the Scherrer constant, $\lambda$ is the x-ray wavelength, $b$ is the full width at half maximum intensity of the film peak in the symmetric $\theta$-$2\theta$ scan, and $\theta$ is the Bragg angle, can be employed to estimate how much of the film has converted to the infinite-layer phase.[39-41] The numerical value of the Scherrer constant $K$ is often approximated to be 0.9,[41] but this value can vary nontrivially upon the geometric factors (i.e. size, shape, and orientation) of the crystallites.[39, 40] Therefore, we determined a suitable value of $K_{film} \approx 1.091$ by comparing the film thickness obtained by measuring XRR on an uncapped $Nd_{0.8}Sr_{0.2}NiO_3$ (001) film to the Scherrer equation. To investigate the generality of $K_{film}$, we compared the thickness values obtained using the Scherrer equation with $K = K_{film}$ to those measured by XRR on other single-crystalline epitaxial perovskite (001) films with varying film thickness ($Nd_{0.8}Sr_{0.2}NiO_3$ (001) and $SrIrO_3$ (001) films), shown in Figure 7(a). Interestingly, the Scherrer thickness values are in good agreement with the XRR thickness values, especially for film thickness below ~ 20 nm. Given that the geometric factors of the perovskite and the infinite-layer structure relevant for the Scherrer constant are similar,[39, 40] this approximation should also be applicable to the infinite-layer phase with reasonable accuracy.

Using this approach, we estimate that the infinite-layer phase of 8.5 nm in thickness (~ 25 unit cells) is stabilized within the capped reduced film shown in Figure 7(b). Although slightly underestimating, this value is in reasonable agreement with the infinite-layer phase thickness of 9.3 nm (~ 27 unit cells) measured from the cross-sectional HAADF-STEM image (Figure 7(b)). This demonstrates that the Scherrer estimate is a useful method for monitoring the crystalline film thickness *ex situ* non-destructively during the reduction process with reasonable accuracy. Furthermore, we note that the infinite-layer phase thickness nearly approaches the maximum possible reduced film thickness $d_{max}$ of 9.7 nm, extracted from the as-grown perovskite film thickness of 10.7 nm (~ 29 unit cells). This corresponds to approximately 2 unit cells of unconverted $Nd_{0.8}Sr_{0.2}NiO_3$ (001), which can be attributed to the interfacial layers as previously observed.[22, 23] In comparison to the partial decomposition of the uncapped film upon reduction (Figure 6(a)), the crystallinity of the film with $SrTiO_3$ capping layer shows significant improvement, with essentially the entire film transformed to the infinite-layer phase.

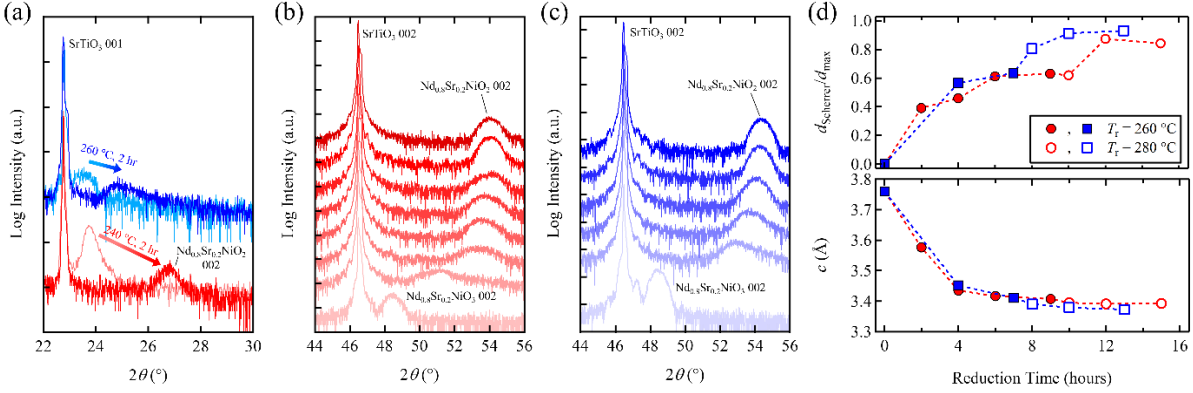

FIG. 8. (a) Shift in the XRD symmetric $\theta$-$2\theta$ scan 001 film peak of a capped sample grown under the high-fluence conditions with ~ 25 nm $SrTiO_3$ (001) capping layer and film thickness of ~ 11 nm after reduction at $T_r = 260$ °C for 2 hours (top), and an uncapped sample (partially optimized sample in Figure 2(a)) with film thickness of ~ 60 nm after reduction at $T_r = 240$ °C for 2 hours (bottom). While the capped sample is still in transition, the uncapped sample is fully reduced to the infinite-layer phase with weaker peak intensity. (b) Evolution of the XRD symmetric $\theta$-$2\theta$ scan around the 002 peak of a capped sample grown under the high-fluence conditions with cap thickness of ~ 25 nm and film thickness of ~ 11 nm, reduced at 260 °C for 9 hours and then 280 °C for 6 hours. Temporal direction is from bottom to top. (c) Evolution of the XRD symmetric $\theta$-$2\theta$ scan around the 002 peak of a nominally similar sample, reduced at 260 °C for 7 hours and then 280 °C for 6 hours. Temporal direction is from bottom to top. (d) Scherrer thickness divided by maximum reduced film thickness (top) and $c$-lattice constant (bottom) plotted against reduction time. The sample in panel (b) is represented as circle markers, and the sample in panel (c) is represented as square markers. $T_r$ is 260 °C for the closed markers and 280 °C for the open markers.

The optimal reduction condition varies as a function of film crystallinity and the thickness of the capping layer, which appears to act as a diffusion barrier to oxygen deintercalation. Highly crystalline samples with ~ 25 nm of $SrTiO_3$ (001) capping layer show gradual XRD peak shifts at $T_r > 240$ °C (Figure 8(a)), while in the limit of no capping layer a complete transition to the infinite-layer phase along with partial film degradation can occur with only 2 hours of reduction at $T_r = 240$ °C (Figure 8(a)). For given crystallinity and capping layer thickness, $T_r$ should be low enough such that the film does not decompose, but also high enough such that the duration of the film exposure to reducing conditions is minimized. As a conservative approach, we performed incremental reductions and assessed the change in the film quality after each increment to minimize the onset of film degradation. Figures 8(b) and 8(c) show two highly crystalline samples with ~ 25 nm of $SrTiO_3$ (001) capping layer under slightly different reduction conditions, where one sample was annealed at $T_r = 260$ °C for 2 more hours than the other. Both samples show saturation in the infinite-layer phase conversion rate after ~ 6 hours of reduction at $T_r = 260$ °C, indicating higher $T_r$ is needed for further reduction (Figure 8(d)). Upon 6 additional hours of reduction at $T_r = 280$ °C, the sample with the shorter overall reduction time is essentially fully reduced, with the Scherrer estimate on the infinite-layer phase conversion rate of 93% corresponding to less than 2 unit cells of unconverted $Nd_{0.8}Sr_{0.2}NiO_3$ (001) (Figure 8(d)). In contrast, the sample with the longer overall reduction time shows the beginning signs of decreasing $d_{Scherrer}/d_{max}$, suggesting the onset of degradation after 6 hours of reduction at $T_r = 280$ °C (Figure 8(d)). On average, 4 - 6 hours of reduction under $T_r \approx 260 - 280$ °C with $SrTiO_3$ capping layer thickness below 25 nm yielded full conversion to the infinite-layer structure.

### III. Comparison of High-Fluence and Low-Fluence Growth Conditions

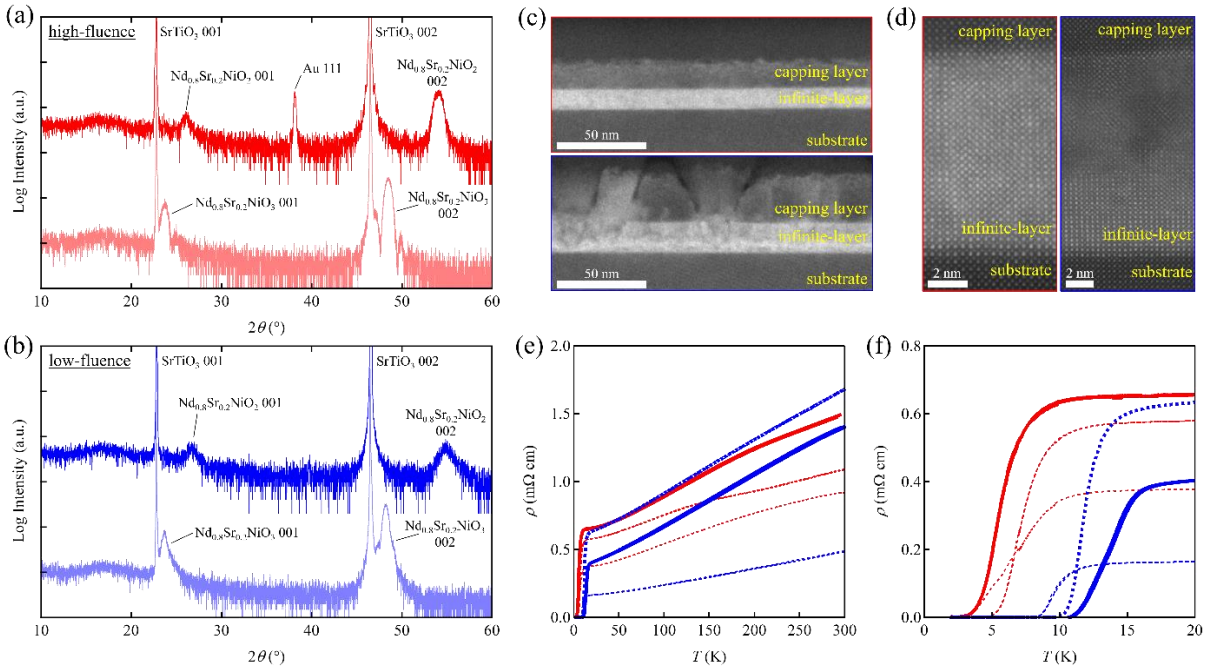

FIG. 9. (a) XRD symmetric $\theta$-$2\theta$ scans of the sample grown under the high-fluence conditions (film thickness ~ 11 nm) as-grown (bottom) and after capping and reduction (9 hours at 260 °C, followed by 3 hours at 280 °C) (top). The Au 111 peak comes from gold contacts evaporated for transport measurements. (b) XRD symmetric $\theta$-$2\theta$ scans of the sample grown under the low-fluence conditions (film thickness ~ 15 nm) as-grown (bottom) and after capping and reduction (4 hours at 280 °C) (top). (c) HAADF STEM image of the reduced samples in panel (a) (top) and panel (b) (bottom). (d) Magnified view of panel (c), with the high-fluence sample image at the left and the low-fluence sample image at the right. (e) $\rho$-$T$ measurement of 6 samples, 3 grown under high-fluence conditions (red) and 3 grown under low-fluence conditions (blue). The thick solid

curves correspond to the samples in panels (a) and (b). (f) Expanded view of panel (e) near the superconducting transition.

With $Nd_{0.8}Sr_{0.2}NiO_3$ nominally optimized in these two different growth conditions, we examine how the difference in the growth conditions affects the crystallinity and the superconducting transition of the resultant $Nd_{0.8}Sr_{0.2}NiO_2$. Figure 9 shows the XRD symmetric $\theta$-$2\theta$ scans of two capped samples: one grown under the high-fluence conditions (Figure 9(a)) and the other grown under the low-fluence conditions (Figure 9(b)). Both samples show prominent 001 perovskite film peaks with no double-peak feature in the 002 film peak, suggesting that the films are dominantly single-phase $Nd_{0.8}Sr_{0.2}NiO_3$ (001) films.

However, we observe multiple signatures indicating that the reduced low-fluence sample has limited crystallinity compared to the high-fluence sample. First, the film peaks of the low-fluence sample are less symmetric and triangular in shape (Figure 9(b)), indicating the presence of nontrivial disorder in the film. Second, upon reduction the low-fluence sample peak intensity decreases (Figure 9(b)), in contrast with the high-fluence sample (Figure 9(a)). This decrease in the peak intensity is also an indication of limited crystallinity in the precursor perovskite phase, resulting in the degradation of film quality during topochemical reduction. This is confirmed from the HAADF STEM images of the two samples after reduction (Figure 9(c)), which show that the infinite-layer region of the low-fluence sample is much less coherent than that of the high-fluence sample. In particular, we observed inclusions and precipitates in the low-fluence sample (Figure 9(d)), similarly to previous reports on partially optimized undoped $NdNiO_3$ thin films.[18] The continuous propagation of disorder into the capping layer again suggests that this disorder originates from the as-grown state before reduction. The higher magnification views of these HAADF STEM images (Figure 9(d)) show that the high-fluence sample displays relatively high crystallinity throughout the entire thickness of the film, while the low-fluence sample maintains crystallinity only near the bottom half of the film. Meanwhile, the low-fluence sample reaches the $c$-lattice constant of 3.34 Å, while the $c$-lattice constant of the high-fluence sample saturates at 3.37 Å; when further reduced under the determined optimal reduction conditions, the high-fluence sample begins to decompose without further decrease in the $c$-lattice constant.

$\rho$-$T$ measurements on the two reduced samples reveal that the low-fluence sample has a higher superconducting transition temperature $T_c$ than the high-fluence sample (Figures 9(e) and 9(f)). For the high-fluence sample shown here, the superconducting transition occurs at an onset of 6.7 K, a midpoint at 5.3 K, and zero resistance at 2.3 K. On the other hand, for the low-fluence sample the superconducting transition occurs at an onset of 15.3 K, a midpoint at 13.3 K, and zero resistance at 10.5 K. The higher superconducting transition temperature for low-fluence samples was reproducibly observed in multiple samples, as shown in Figures 9(e) and 9(f).

There are some observations worth discussing at this point. First, the wide sample-to-sample variation in $T_c$ in the first report[15] has now been reproducibly narrowed in this study, controlled in part by the use of precise imaging conditions for ablation. For the two perovskite phase growth conditions optimized and studied here, the high-fluence samples have significantly better crystallinity in the reduced phase. Therefore, it is somewhat surprising that the low-fluence samples show systematically higher $T_c$. While the origin of this distinction is yet unclear, we note the difference in the $c$-lattice constant of the two groups of samples, which may indicate that the distance between Ni-O planes is highly relevant for $T_c$. On the other hand, for further systematic studies on superconductivity and normal state properties, high-fluence conditions may be preferable given the more uniform crystallinity.

**Conclusion**

In summary, we have investigated the synthesis of infinite-layer nickelate $Nd_{0.8}Sr_{0.2}NiO_2$ (001) epitaxial thin films. The two principal technical issues we identified were the stabilization of the doped perovskite phase, and the balance between complete topotactic reduction versus subsequent decomposition.

We emphasize that the current conditions presented may not be the global optimum for PLD growth, given the many parameters and potentially competing factors for synthesis of the perovskite phase and the reduction to the infinite-layer phase. Nevertheless, we hope the current work will be valuable to the community interested in this system. We further note that high-quality perovskite nickelate films have also been synthesized by other techniques, such as molecular beam epitaxy[18, 42-45] and sputtering.[28, 46-48] It will be intriguing to see if these techniques provide new opportunities in the synthesis of these compounds.

There clearly remain many open questions on the effect of the growth conditions on the structural properties and chemical composition of these infinite-layer nickelates, particularly on the microscopic scale. While the wide sample-to-sample variation of the superconducting $T_c$ initially reported[15] has now been significantly narrowed, the key factors for the variation in $T_c$ for the same nominal composition remains unknown. Further investigation of this aspect may provide insight into some of the critical factors governing superconductivity. Further studies including comprehensive local crystallographic characterization, high-resolution compositional analysis, the effect of substrate strain, spectroscopic measurements, and doping-dependence investigations are needed to further understand the materials factors relevant for the occurrence of superconductivity.

**Acknowledgments**

This work was supported by the US Department of Energy, Office of Basic Energy Sciences, Division of Materials Sciences and Engineering, under contract number DE-AC02-76SF00515. D.L. acknowledges partial support by the Swiss National Science Foundation (P2GEP2_168277), and the Gordon and Betty Moore Foundation's Emergent Phenomena in Quantum Systems Initiative through grant number GBMF4415. B.H.G. and L.F.K. acknowledge support by the Department of Defense Air Force Office of Scientific Research (No. FA 9550-16-1-0305) and the Packard Foundation. This work made use of a Helios FIB supported by NSF (DMR-1539918) and the Cornell Center for Materials Research Shared Facilities which are supported through the NSF MRSEC program (DMR-1719875). The FEI Titan Themis 300 was acquired through No. NSF-MRI-1429155, with additional support from Cornell University, the Weill Institute, and the Kavli Institute at Cornell.

# References

1. M. Crespin, P. Levitz and L. Gatineau, J. Chem. Soc., Faraday Trans. 2 **79**, 1181 (1983).
2. P. Levitz, M. Crespin and L. Gatineau, J. Chem. Soc., Faraday Trans. 2 **79**, 1195 (1983).
3. P. Lacorre, J. Solid State Chem. **97**, 495 (1992).
4. M. A. Hayward, M. A. Green, M. J. Rosseinsky and J. Sloan, J. Am. Chem. Soc. **121**, 8843 (1999).
5. M. A. Hayward and M. J. Rosseinsky, Sol. State Sci. **5**, 839 (2003).
6. V. V. Poltavets, K. A. Lokshin, S. Dikmen, M. Croft, T. Egami and M. Greenblatt, J. Am. Chem. Soc. **128**, 9050 (2006).
7. J. Zhang, A. S. Botana, J. W. Freeland, D. Phelan, H. Zheng, V. Pardo, M. R. Norman and J. F. Mitchell, Nat. Phys. **13**, 864 (2017).
8. T. Siegrist, S. M. Zahurak, D. W. Murphy and R. S. Roth, Nature **334**, 231 (1988).
9. M. G. Smith, A. Manthiram, J. Zhou, J. B. Goodenough and J. T. Markert, Nature **351**, 549 (1991).
10. M. Azuma, Z. Hiroi, M. Takano, Y. Bando and Y. Takeda, Nature **356**, 775 (1992).
11. V. I. Anisimov, D. Bukhvalov and T. M. Rice, Phys. Rev. B **59**, 7901 (1999).
12. K.-W. Lee and W. E. Pickett, Phys. Rev. B **70**, 165109 (2004).
13. M. Kawai, S. Inoue, M. Mizumaki, N. Kawamura, N. Ichikawa and Y. Shimakawa, Appl. Phys. Lett. **94**, 082102 (2009).
14. D. Kaneko, K. Yamagishi, A. Tsukada, T. Manabe and M. Naito, Physica C **469**, 936 (2009).
15. D. Li, K. Lee, B. Y. Wang, M. Osada, S. Crossley, H. R. Lee, Y. Cui, Y. Hikita and H. Y. Hwang, Nature **572**, 624 (2019).
16. G. Catalan, R. M. Bowman and J. M. Gregg, J. Appl. Phys. **87**, 606 (1999).
17. E. Breckenfeld, Z. Chen, A. R. Damodaran and L. W. Martin, ACS Appl. Mater. Interfaces **6**, 22436 (2014).
18. A. J. Hauser, E. Mikheev, N. E. Moreno, J. Hwang, J. Y. Zhang and S. Stemmer, Appl. Phys. Lett. **106**, 092104 (2015).
19. D. Preziosi, A. Sander, A. Barthélémy and M. Bibes, AIP Adv. **7**, 015210 (2017).
20. G. Demazeau, A. Marbeuf, M. Pouchard and P. Hagenmuller, J. Solid State Chem. **3**, 582 (1971).
21. M. Kawai, K. Matsumoto, N. Ichikawa, M. Mizumaki, O. Sakata, N. Kawamura, S. Kimura and Y. Shimakawa, Cryst. Growth Des. **10**, 2044 (2010).
22. T. Onozuka, A. Chikamatsu, T. Katayama, T. Fukumura and T. Hasegawa, Dalton Trans. **45**, 12114 (2016).
23. A. Ikeda, Y. Krockenberger, H. Irie, M. Naito and H. Yamamoto, Appl. Phys. Express **9**, 061101 (2016).
24. K. Lee, M. Osada, H. Y. Hwang and Y. Hikita, J. Phys. Chem. Lett. **10**, 1516 (2019).
25. J. Liu, M. Kareev, B. Gray, J. W. Kim, P. Ryan, B. Dabrowski, J. W. Freeland and J. Chakhalian, Appl. Phys. Lett. **96**, 233110 (2010).
26. D. Preziosi, L. Lopez-Mir, X. Li, T. Cornelissen, J. H. Lee, F. Trier, K. Bouzehouane, S. Valencia, A. Gloter, A. Barthélémy and M. Bibes, Nano Lett. **18**, 2226 (2018).
27. P. Lacorre, J. B. Torrance, J. Pannetier, A. I. Nazzal, P. W. Wang and T. C. Huang, J. Solid State Chem. **91**, 225 (1991).
28. R. Scherwitzl, P. Zubko, I. G. Lezama, S. Ono, A. F. Morpurgo, G. Catalan and J. M. Triscone, Adv. Mater. **22**, 5517 (2010).

29. P. A. W. v. d. Heide, Surf. Interface Anal. **33**, 414 (2002).
30. M. Miura, S. Adachi, T. Shimode, K. Wada, A. Takemori, N. Chikumoto, K. Nakao and K. Tanabe, Appl. Phys. Express **6**, 093101 (2013).
31. K. Iida, V. Grinenko, F. Kurth, A. Ichinose, I. Tsukada, E. Ahrens, A. Pukenas, P. Chekhonin, W. Skrotzki, A. Teresiak, R. Hühne, S. Aswartham, S. Wurmehl, I. Mönch, M. Erbe, J. Hänisch, B. Holzapfel, S.-L. Drechsler and D. V. Efremov, Sci. Rep. **6**, 28390 (2016).
32. J. Y. Koo, H. Kwon, M. Ahn, M. Choi, J.-W. Son, J. W. Han and W. Lee, ACS Appl. Mater. Interfaces **10**, 8057 (2018).
33. T. Sarkar, R. L. Greene and S. D. Sarma, Phys. Rev. B **98**, 224503 (2018).
34. Q. Guo, S. Farokhipoor, C. Magén, F. Rivadulla and B. Noheda, arXiv:1909.06256 [cond-mat.str-el] (2019).
35. Y. Tokuda, S. Kobayashi, T. Ohnishi, T. Mizoguchi, N. Shibata, Y. Ikuhara and T. Yamamoto, Appl. Phys. Lett. **99**, 173109 (2011).
36. J. Bak, H. B. Bae, J. Kim, J. Oh and S.-Y. Chung, Nano Lett. **17**, 3126 (2017).
37. A. Olafsen, H. Fjellvåg and B. C. Hauback, J. Solid State Chem. **151**, 46 (2000).
38. P. F. Fewster, Rep. Prog. Phys. **59**, 1339 (1996).
39. H. P. Klug and L. E. Alexander, *X-Ray Diffraction Procedures*, 2 ed. (Wiley, New York, 1974).
40. J. I. Langford and A. J. C. Wilson, J. Appl. Cryst. **11**, 102 (1978).
41. U. Holzwarth and N. Gibson, Nat. Nanotechnol. **6**, 534 (2011).
42. J. F. DeNatale and P. H. Kobrin, J. Mater. Res. **10**, 2992 (2011).
43. P. D. C. King, H. I. Wei, Y. F. Nie, M. Uchida, C. Adamo, S. Zhu, X. He, I. Božović, D. G. Schlom and K. M. Shen, Nat. Nanotechnol. **9**, 443 (2014).
44. F. Wrobel, A. F. Mark, G. Christiani, W. Sigle, H.-U. Habermeier, P. A. van Aken, G. Logvenov, B. Keimer and E. Benckiser, Appl. Phys. Lett. **110**, 041606 (2017).
45. A. S. Disa, A. B. Georgescu, J. L. Hart, D. P. Kumah, P. Shafer, E. Arenholz, D. A. Arena, S. Ismail-Beigi, M. L. Taheri and F. J. Walker, Phys. Rev. Mater. **1**, 024410 (2017).
46. C. C. Yang, M. S. Chen, T. J. Hong, C. M. Wu, J. M. Wu and T. B. Wu, Appl. Phys. Lett. **66**, 2643 (1995).
47. N. Wakiya, T. Azuma, K. Shinozaki and N. Mizutani, Thin Solid Films **410**, 114 (2002).
48. R. Scherwitzl, P. Zubko, C. Lichtensteiger and J. M. Triscone, Appl. Phys. Lett. **95**, 222114 (2009).